\documentclass[a4paper,11pt]{article}
\usepackage{preamble/jheppubMOD} 
\usepackage{lipsum}
\usepackage{graphicx}  
\usepackage{multirow}
\usepackage[x11names,dvipsnames]{xcolor}

\usepackage{preamble/mciteplus}

\hypersetup{
    colorlinks,
    linkcolor={jpac-blue},
    citecolor={jpac-blue},
    urlcolor={jpac-blue}
}

\usepackage{lmodern}
\usepackage{amsmath,amssymb}
\usepackage{mathtools}
\usepackage[numbers,sort&compress]{natbib}

\usepackage{tikz}
\usepackage{tikz-3dplot}
\usepackage{pgfplots,pgfplotstable}
\usetikzlibrary{pgfplots.groupplots}
\usetikzlibrary{decorations.markings}
\usetikzlibrary{shapes}
\usetikzlibrary{shapes.misc}
\usetikzlibrary{patterns}
\usepgfplotslibrary{fillbetween}
\pgfplotsset{compat=newest}
\usetikzlibrary{positioning}
\usetikzlibrary{calc}

\usepackage{lipsum}
\usepackage{stackrel}

\usepackage[compat=1.1.0]{tikz-feynman}

\usepackage{orcidlink}

\usepackage{cleveref}
\usepackage{ifthen}
\usepackage{xspace}
\usepackage[shortlabels]{enumitem}

\definecolor{jpac-blue}{RGB}{ 31,119,180}
\definecolor{jpac-red}{RGB}{214,39, 40}
\definecolor{jpac-green}{RGB}{ 44,160, 44}
\definecolor{jpac-orange}{RGB}{255,127, 14}
\definecolor{jpac-purple}{RGB}{148,103,189}
\definecolor{jpac-brown}{RGB}{140, 86, 75}
\definecolor{jpac-pink}{RGB}{227,119,194}
\definecolor{jpac-gold}{RGB}{188,189, 34}
\definecolor{jpac-aqua}{RGB}{ 23,190,207}
\definecolor{jpac-grey}{RGB}{127,127,127}

\pgfdeclarepatternformonly{crosshatchD}{\pgfqpoint{-1.5pt}{-1.5pt}}{\pgfqpoint{4pt}{4pt}}{\pgfqpoint{2.5pt}{2.5pt}}%
{%
  \pgfsetlinewidth{0.4pt}%
  \pgfpathmoveto{\pgfqpoint{3.1pt}{0pt}}%
  \pgfpathlineto{\pgfqpoint{0pt}{3.1pt}}%
  \pgfpathmoveto{\pgfqpoint{0pt}{0pt}}%
  \pgfpathlineto{\pgfqpoint{3.1pt}{3.1pt}}%
  \pgfusepath{stroke}%
}%

\tikzset{vpiFF/.style={draw=jpac-red,fill=jpac-pink!40 ,circle,very thick,inner sep=4.5pt}}
\tikzset{vfone/.style={draw=jpac-blue,fill=jpac-aqua!40,circle,very thick,inner sep=4.5pt}}

\tikzfeynmanset{pion/.style={black,charged scalar, dash pattern=on 3pt off 2pt,very thick}}
\tikzfeynmanset{phot/.style={black,photon,very thick}}
\tikzfeynmanset{omeg/.style={black,fermion,very thick}}
\tikzfeynmanset{cutAs/.style={black!80,very thick}}
\tikzfeynmanset{cutBs/.style={black!80,thin,decoration={pre length=0.06cm, post length=0.0cm,border,segment length=0.4mm,amplitude=1mm,angle=270},
            postaction={decorate,draw}}}

\pgfplotsset{error bar legend/.style={%
    /pgfplots/legend image code/.prefix code={%
      \pgfkeysgetvalue{/pgfplots/error bars/error mark}{\pgfplotserrorbarsmark}%
      \draw[%
        /pgfplots/every error bar, 
        mark=\pgfplotserrorbarsmark, 
        /pgfplots/error bars/error mark options, 
        sharp plot,
        ##1,
      ] plot coordinates {(0.30cm, -0.15cm) (0.30cm, 0.15cm)};%
    }
  }
}

\pgfplotsset{DataNA6009/.style={line width=1pt,color=jpac-orange,only marks, mark=*,mark size=1.35pt,opacity=0.65, fill opacity=0.75,on layer={axis background},error bar legend,error bars/.cd, y dir=both,y explicit,error bar style={opacity=0.75,fill opacity=0.75,line width=0.75pt}}}

\pgfplotsset{DataNA6016/.style={line width=1pt,color=jpac-green,only marks, mark=triangle*,mark size=1.35pt,opacity=0.65, fill opacity=0.75,on layer={axis background},error bar legend,error bars/.cd, y dir=both,y explicit,error bar style={opacity=0.75,fill opacity=0.75,line width=0.75pt}}}

\pgfplotsset{DataMAMI/.style={line width=1pt,color=jpac-purple,only marks, mark=square*,mark size=1.35pt,opacity=0.65, fill opacity=0.75,on layer={axis background},error bar legend,error bars/.cd, y dir=both,y explicit,error bar style={opacity=0.75,fill opacity=0.75,line width=0.75pt}}}

\pgfplotsset{DataSND11/.style={line width=1pt,color=jpac-red,only marks, mark=o,mark size=1.35pt,opacity=0.65, fill opacity=0.75,on layer={axis background},error bar legend,error bars/.cd, y dir=both,y explicit,error bar style={opacity=0.75,fill opacity=0.75,line width=0.75pt}}}

\pgfplotsset{DataSND13/.style={line width=1pt,color=jpac-blue,only marks, mark=triangle,mark size=1.75pt,opacity=0.65, fill opacity=0.75,on layer={axis background},error bar legend,error bars/.cd, y dir=both,y explicit,error bar style={opacity=0.75,fill opacity=0.75,line width=0.75pt}}}

\pgfplotsset{DataCMD03/.style={line width=1pt,color=yellow,only marks, mark=square,mark size=1.35pt,opacity=0.65, fill opacity=0.75,on layer={axis background},error bar legend,error bars/.cd, y dir=both,y explicit,error bar style={opacity=0.75,fill opacity=0.75,line width=0.75pt}}}

\colorlet{lowphi}{jpac-blue}
\colorlet{higphi}{jpac-red}
\colorlet{expcolor}{jpac-green}

\colorlet{KTzero}{jpac-green}
\pgfplotsset{KTzerolA/.style={thick,color=KTzero,no markers}}
\pgfplotsset{KTzerolB/.style={thick,color=KTzero,no markers,dash pattern=on 3pt off 2pt}}

\colorlet{KTones}{jpac-blue}
\pgfplotsset{KToneslA/.style={thick,color=KTones,no markers}}
\pgfplotsset{KToneslB/.style={thick,color=KTones,no markers,dash pattern=on 3pt off 2pt}}

\pgfplotsset{HistoStyleL/.style={const plot,draw=none,fill=lowphi,opacity=0.5}}
\pgfplotsset{HistoStyleH/.style={const plot,draw=none,fill=higphi,opacity=0.5}}

\newcommand\bsub{\begin{subequations}}
\newcommand\esub{\end{subequations}}

\newcommand{\eg}{{\it e.g.}\xspace}
\newcommand{\ie}{{\it i.e.}\xspace}
\newcommand{\cf}{{\it cf.}\xspace}

\newcommand{\kev}{\ensuremath{{\mathrm{\,ke\kern -0.1em V}}}\xspace}
\newcommand{\mev}{\ensuremath{{\mathrm{\,Me\kern -0.1em V}}}\xspace}
\newcommand{\gev}{\ensuremath{{\mathrm{\,Ge\kern -0.1em V}}}\xspace}
\newcommand{\gevsq}{\ensuremath{{\mathrm{\,Ge\kern -0.1em V}^2}}\xspace}
\newcommand{\tev}{\ensuremath{{\mathrm{\,Te\kern -0.1em V}}}\xspace}

\def\XXint#1#2#3{{\setbox0=\hbox{$#1{#2#3}{\int}$}\vcenter{\hbox{$#2#3$}}\kern-.5\wd0}}

\begin{document}

\title{\boldmath Khuri-Treiman analysis of $J/\psi\to\pi^{+}\pi^{-}\pi^{0}$}

\collaboration{JPAC Collaboration}
\collaborationImg{\includegraphics[height=2cm,keepaspectratio]{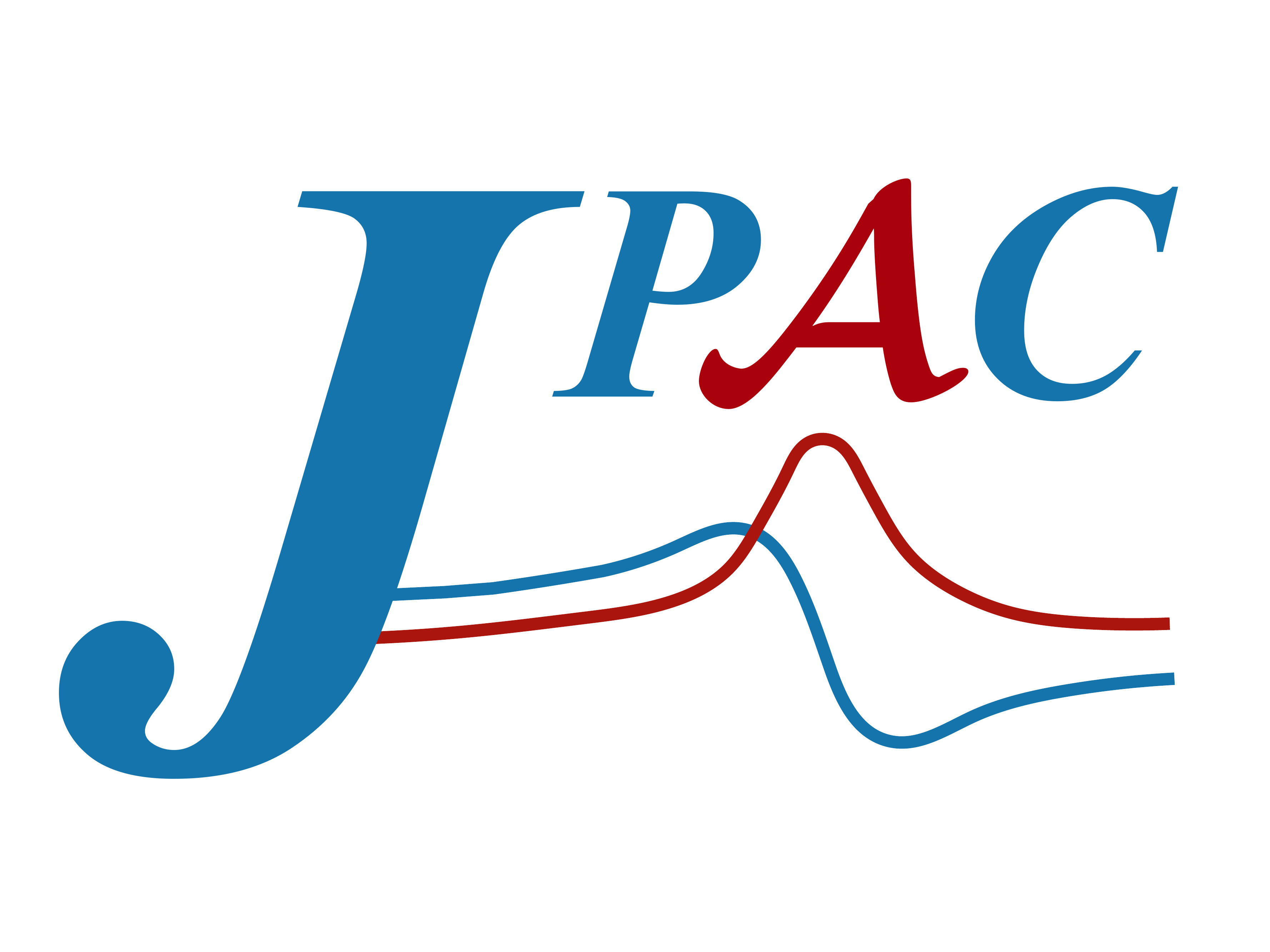}}
\preprint{LA-UR-23-23576, JLAB-THY-23-3796}

\newcommand{\ceem}{Center for  Exploration  of  Energy  and  Matter,
Indiana  University,
Bloomington,  IN  47403,  USA}
\newcommand{\indiana}{Department of Physics,
Indiana  University,
Bloomington,  IN  47405,  USA}
\newcommand{\jlab}{Theory Center,
Thomas  Jefferson  National  Accelerator  Facility,
Newport  News,  VA  23606,  USA}
\newcommand{\icn}{Instituto de Ciencias Nucleares, 
Universidad Nacional Aut\'onoma de M\'exico, Ciudad de M\'exico 04510, Mexico}
\newcommand{\uned}{Departamento de F\'isica Interdisciplinar, Universidad Nacional de Educaci\'on a Distancia (UNED), Madrid E-28040, Spain}
\newcommand{\hiskp}{Universit\"at Bonn,
Helmholtz-Institut f\"ur Strahlen- und Kernphysik, 53115 Bonn, Germany}
\newcommand{\origins}{ORIGINS Excellence Cluster, 85748 Garching, Germany}
\newcommand{\lmu}{Ludwig-Maximilians-Universit{\"a}t, 80539 Munich, Germany}
\newcommand{\ect}{European Centre for Theoretical Studies in Nuclear Physics and related Areas (ECT$^*$) and Fondazione Bruno Kessler, Villazzano (Trento), I-38123, Italy}
\newcommand{\genova}{INFN Sezione di Genova, Genova, I-16146, Italy}
\newcommand{\cern}{CERN, 1211 Geneva 23, Switzerland}
\newcommand{\ucm}{Departamento de F\'isica Te\'orica, Universidad Complutense de Madrid and IPARCOS, 28040 Madrid, Spain}
\newcommand{\mainz}{Institut f\"ur Kernphysik \& PRISMA$^+$  Cluster of Excellence, Johannes Gutenberg Universit\"at,  D-55099 Mainz, Germany}
\newcommand{\lanl}{Theoretical Division, Los Alamos National Laboratory, Los Alamos, NM 87545, USA}
\newcommand{\ific}{Instituto de F\'isica Corpuscular (IFIC), Centro Mixto CSIC-Universidad de Valencia, E-46071 Valencia, Spain}
\newcommand{\ub}{Departament de F\'isica Qu\`antica i Astrof\'isica and Institut de Ci\`encies del Cosmos, Universitat de Barcelona, E08028, Spain}
\newcommand{\ifj}{Institute of Nuclear Physics, Polish Academy of Sciences, 31-342 Kraków, Poland}
\newcommand{\catania}{INFN Sezione di Catania, I-95123 Catania, Italy}
\newcommand{\messina}{Dipartimento di Scienze Matematiche e Informatiche, Scienze Fisiche e Scienze della Terra,
Universit\`a degli Studi di Messina, I-98122 Messina, Italy}
\newcommand{\odu}{Department of Physics, Old Dominion University, Norfolk, VA 23529, USA}
\newcommand{\scnuIQM}{Guangdong Provincial Key Laboratory of Nuclear Science, Institute of Quantum Matter, South China Normal University, Guangzhou 510006, China}
\newcommand{\scnuJLQM}{Guangdong-Hong Kong Joint Laboratory of Quantum Matter, Southern Nuclear Science Computing Center, South China Normal University, Guangzhou 510006, China}
\newcommand{\AGH}{AGH University of Krakow, Faculty of Physics and Applied Computer Science, al. Adama Mickiewicza 30
30-059 Krak\'ow}

\author[a]{M.~Albaladejo\orcidlink{0000-0001-7340-9235}}
\emailAdd{Miguel.Albaladejo@ific.uv.es}
\author[b]{S.~Gonz\`{a}lez-Sol\'is\orcidlink{0000-0003-1947-5420}}
\emailAdd{sergig@lanl.gov}
\author[c]{\L. Bibrzycki\orcidlink{0000-0002-6117-4894}}
\author[d,e]{C.~Fern\'andez-Ram\'irez\orcidlink{0000-0001-8979-5660}}
\author[f]{N.~Hammoud\orcidlink{0000-0002-8395-0647}}
\author[g,h]{V.~Mathieu\orcidlink{0000-0003-4955-3311}}
\author[i,j]{M.~Mikhasenko\orcidlink{0000-0002-6969-2063}}
\author[k]{G.~Monta\~na\orcidlink{0000-0001-8093-6682}}
\author[g]{R.~J.~Perry\orcidlink{0000-0002-2954-5050}}
\author[l,m]{A.~Pilloni\orcidlink{0000-0003-4257-0928}}
\author[k]{A.~Rodas\orcidlink{0000-0003-2702-5286}}
\author[n,o]{W.~A.~Smith\orcidlink{0009-0001-3244-6889}}
\author[k,n,o]{A.~Szczepaniak\orcidlink{0000-0002-4156-5492}}
\author[p,q]{D.~Winney\orcidlink{0000-0002-8076-243X}}

\affiliation[a]{\ific}
\affiliation[b]{\lanl}
\affiliation[c]{\AGH}
\affiliation[d]{\uned}
\affiliation[e]{\icn}
\affiliation[f]{\ifj}
\affiliation[g]{\ub}
\affiliation[h]{\ucm}
\affiliation[i]{\origins}
\affiliation[j]{\lmu}
\affiliation[k]{\jlab}
\affiliation[l]{\messina}
\affiliation[m]{\catania}
\affiliation[n]{\indiana}
\affiliation[o]{\ceem}
\affiliation[p]{\scnuIQM}
\affiliation[q]{\scnuJLQM}

\abstract{
We study the decay $J/\psi\to\pi^{+}\pi^{-}\pi^{0}$ within the framework of the Khuri-Treiman equations.
We find that the BESIII experimental di-pion mass distribution in the $\rho(770)$-region is well reproduced with a once-subtracted $P$-wave amplitude. 
Furthermore, we show that $F$-wave contributions to the amplitude improve the description of the data in the $\pi\pi$ mass region around 1.5 GeV.
We also present predictions for the $J/\psi\to\pi^{0}\gamma^{*}$ transition form factor.
}

\frenchspacing
\toccontinuoustrue
\maketitle

\section{Introduction}\label{sec:introduction}

Decays of the lowest-lying charmonium states provide an excellent environment to study light hadron spectroscopy, search for exotic mesons, test QCD and QCD-based models, as well as testing theoretical techniques in a region where both non-perturbative and perturbative QCD effects play a role.

In this work we analyze the decay $J/\psi\to\pi^{+}\pi^{-}\pi^{0}$, to study the dynamics of the three-pion system at low and intermediate energies under rather clean conditions.
Here, the final state invariant mass distribution can contain contributions from the $P$-wave $(J^{PC}=1^{--})$ and $F$-wave $(J^{PC}=3^{--})$ states of the $\pi\pi$ subsystem. 
Previous experimental studies from BESII~\cite{BES:2004mxa} and BABAR~\cite{BaBar:2004ytv} showed that the $P$-wave $\rho(770)\pi$ intermediate state dominates the process, but limited statistics prevented any detailed study of substructures in the $3\pi$ system.
While the dominance of the $\rho(770)$ resonance can be clearly seen in the Dalitz plot distribution and projection measurements by the BESIII collaboration obtained with roughly 1.9 million $J/\psi\to\pi^{+}\pi^{-}\pi^{0}$ events~\cite{BESIII:2012vmy}, there are hints of contributions other than the $\rho(770)$.
For example, the absence of events in the center of the Dalitz plot indicates the contribution from additional states and/or partial waves which may interfere destructively with the $\rho(770)$.
Exactly the opposite situation is found for the partner reaction $\psi(2S)\to\pi^{+}\pi^{-}\pi^{0}$.
There, the 7872 events from BESIII~\cite{BESIII:2012vmy} show a completely different shape of the $\pi\pi$ invariant mass distribution and the Dalitz plot --- the $\rho\pi$ contribution is subleading and almost all events are found in the center of the Dalitz plot, with data indicating that the main contribution comes from a higher mass resonance, {\it i.e.} the $\rho(2150)$ resonance with $J^{PC}=1^{--}$.
The different picture between the $J/\psi$ and $\psi(2S)$ decays into $\pi^{+}\pi^{-}\pi^{0}$
and the lack of reasonable explanations within the quark model is known as the $\rho\pi$ puzzle and still remains largely unresolved (see \eg Refs.\,\cite{Chen:1998ma,Mo:2006cy,Wang:2012mf,Kivel:2023fgu,Yan:2023nqz}, and references therein).
New high-statistics BESIII data on $J/\psi$ decays will soon be available \cite{BESIII:2021cxx,BESIII:2020nme}, which could be used to greatly improve the theoretical uncertainties associated to vector charmonium decays. 
In particular, they might help clarify the $\rho\pi$ puzzle, 
as well as provide access to high-precision $\rho$-$\omega$ mixing effect analyses and motivate coupled channel studies with the decays $J/\psi\to K^{+}K^{-}\pi^{0}$ and $J/\psi\to K_{S}K^{\pm}\pi^{\mp}$.

The decay $J/\psi\to\pi^{+}\pi^{-}\pi^{0}$ has previously been studied within the context of the Veneziano model~\cite{Szczepaniak:2014bsa}, and using aspects of unitarity and analyticity constraints~\cite{Guo:2010gx,Guo:2011aa}.
Here, we adapt the Khuri-Treiman (KT) framework~\cite{Khuri:1960zz}, applied extensively in the isospin-violating decay $\eta\to3\pi$~\cite{Anisovich:1996tx,Guo:2014vya,Guo:2015zqa,Colangelo:2016jmc,Colangelo:2018jxw, Albaladejo:2017hhj,Gasser:2018qtg} and in the decay of light vector isoscalar resonances $\omega,\phi\to3\pi$~\cite{Niecknig:2012sj,Danilkin:2014cra,JPAC:2020umo}, to the analysis of the vector charmonium decay $J/\psi\to\pi^{+}\pi^{-}\pi^{0}$.
We show that one subtraction in the KT equations satisfactorily describes the BESIII experimental di-pion mass distribution at the peak of the $\rho(770)$.
In addition, we find that $F$-wave effects are needed to describe the intermediate energy region around 1.5 GeV.
We also apply our analysis techniques to predict the $J/\psi\to\pi^{0}\gamma^{*}$ transition form factor.
Our study lays the groundwork for a detailed analysis of $J/\psi$ decays using the large data sample currently being collected at BESIII.

This paper is organized as follows. 
In Section~\ref{sec:formalism} we review the KT formalism for the $J/\psi\to 3\pi$ decay.
In Section~\ref{sec:results} we apply the formalism to the BESIII data and discuss the results. 
In Section~\ref{sec:TFF}, we present predictions for the $J/\psi\to\pi^{0}\gamma^{*}$ transition form factor, and we summarize our findings in Section~\ref{sec:conclusions}. 

\section{Formalism}\label{sec:formalism}

\subsection{Decay amplitude and kinematics}\label{subsec:kin}

The amplitude for the decay $J/\psi(p_{V})\to\pi^0(p_0) \; \pi^+(p_+) \; \pi^-(p_-)$ can be expressed in terms of a kinematic prefactor and a single invariant scalar function $F(s,t,u)$ containing the dynamical information,
\begin{equation}
\mathcal{M}(s,t,u)=i\,\epsilon_{\mu\nu\alpha\beta}\;\epsilon^{\mu}(p_{V})\,p_{+}^{\nu}\,p_{-}^{\alpha}\,p_{0}^{\beta}\,\,F(s,t,u)\,,
\label{Eq:AmplitudeOmega3Pi}
\end{equation}
where $\epsilon_{\mu\nu\alpha\beta}$ is the Levi-Civita tensor and $\epsilon^{\mu}(p_{V})$ is the polarization vector of the $J/\psi$ meson. 
The particle momenta are related to the Mandelstam variables through:
\begin{equation}
    s  = (p_+ + p_-)^2\,,\quad t = (p_0 + p_+)^2\,,\quad u  = (p_0 + p_-)^2\,,
\end{equation}
with $s+t+u=m_{J/\psi}^{2}+3m_{\pi}^{2}$. In this paper, we work in the isospin limit with $m_{\pi}\doteq m_{\pi^{\pm}}=m_{\pi^{0}}$ and $m_{\pi}=(2m_{\pi^{\pm}}+m_{\pi^{0}})/3$. The scattering angle in the $s$-channel, defined by the center of mass of the $\pi^+\pi^-$ pair, is denoted by $\theta_s$ and is given by:
\begin{equation}\label{Eq:CosTheta}
    \cos\theta_s(s,t,u) = \frac{t - u}{4 \, p(s) \,q(s)}\,,\quad \sin\theta_s(s,t,u)=\frac{\sqrt{\phi(s,t,u)}}{2\sqrt{s}\,p(s) \, q(s)}\,,
\end{equation}
where the momenta $p(s)$ and $q(s)$,
\begin{equation}
   p(s)= \frac{\lambda^{\frac{1}{2}}(s,m_{\pi}^2,m_{\pi}^2)}{2\sqrt{s}}\,,\quad q(s) = \frac{\lambda^{\frac{1}{2}}(s,m_{J/\psi}^2,m_{\pi}^2)}{2\sqrt{s}}\,,
\end{equation}
are, respectively, the momenta of the $\pi^\pm$ and $\pi^0$ in the $s$-channel. 
$\lambda(a,b,c) = a^2 + b^2 + c^2 - 2ab-2bc-2ca$ is the 
K\"all\'en, or triangle, function~\cite{Kallen:1964lxa}.
The zeroes of the well-known Kibble function~\cite{Kibble:1960zz} ,
\begin{equation}
    \phi(s,t,u)=\left(2\sqrt{s} \; \sin\theta_{s} \; p(s) \, q(s)\right)^{2}=s\,t\,u - m_{\pi}^2 (m_{J/\psi}^2 - m_{\pi}^2)^2~,
    \label{Eq:KibbleFunction}
\end{equation}
define the boundaries of the physical regions of the process.
The Dalitz-plot boundaries in $t$ for a given value of $s$ for $J/\psi\to3\pi$ lie within the interval $[t_{\rm{min}}(s), \; t_{\rm{max}}(s)]$, with 
\begin{equation}
    t_{\rm{max,min}}(s)= \frac{m_{J/\psi}^{2}+3m_{\pi}^{2}-s}{2} \pm  2 \, p(s) \, q(s)~,
\label{Eq:tmaxmin}
\end{equation}
while the allowed range for $s$ is given by $s_{\rm{min}}=4m_{\pi}^{2}$ to $s_{\rm{max}}=(m_{J/\psi}-m_{\pi})^{2}\,.$

Finally, the measured differential decay width can be written in terms of the invariant amplitude $F(s,t,u)$ as
\begin{equation}
\frac{d^2 \Gamma}{d s\,d t} = \frac{1}{(2\pi)^3}\, \frac{1}{32\,m_{J/\psi}^3} \, \frac{1}{3} \frac{\phi(s,t,u)}{4}  \; | F(s,t,u) |^2\,.
\label{Eq:DecayWidth}
\end{equation}

\subsection[Khuri--Treiman equations for $J/\psi \to 3\pi$]{\boldmath Khuri--Treiman equations for $J/\psi \to 3\pi$}\label{subsec:KT}

The KT formalism for the $J/\psi \to 3\pi$ amplitude $F(s,t,u)$ is formally identical to the well-established one for the $\omega\to3\pi$ decay amplitude~\cite{Niecknig:2012sj,Danilkin:2014cra,JPAC:2020umo,JPAC:2021rxu}, and has been discussed in Ref.\,\cite{Kubis:2014gka} (see also Ref.\,\cite{Stamen:2022eda}). 
As shown in these references, the $s$-channel partial-wave expansion for $F(s,t,u)$ is given by
\begin{equation}
F(s,t,u)=\sum_{J\,\rm{odd}}^{\infty}(p(s) \, q(s))^{J-1} \; P_{J}^{\prime}(z_{s}) \; f_{J}(s)\,,
\label{Eq:AmplitudeF}
\end{equation}
where $z_{s}=\cos\theta_{s}$ and $P_{J}^{\prime}(z_{s})$ is the derivative of the Legendre polynomial.
 The KT representation of the scalar function $F(s,t,u)$ in Eq.~\eqref{Eq:AmplitudeF} may be obtained by replacing the infinite sum of partial waves in the $s$-channel with the sum of three so-called isobar amplitudes, one for each of the $s$-, $t$- and $u$-channels. 
 By truncating the partial wave expansion of each isobar amplitude at $J_\text{max}=1$ we obtain the following crossing-symmetric isobar decomposition~\cite{Niecknig:2012sj,Danilkin:2014cra,Albaladejo:2019huw}:
\begin{equation}
F(s,t,u)=F_{1}(s)+F_{1}(t)+F_{1}(u)\,, 
\label{Eq:KTdecomposition}
\end{equation}
where each isobar amplitude, $F_{1}(x)$, has only a right-hand or unitary cut in its respective Mandelstam variable. 
The  relation between $F_{1}(s)$ and $f_1(s)$ is obtained by projecting Eq.~\eqref{Eq:KTdecomposition} onto the $s$-channel partial wave,
\begin{align}
\label{Eq:f1}
f_1(s)& = F_{1}(s)+\hat{F}_{1}(s)\,, \\[1ex]
\hat{F}_{1}(s)& \equiv 3\int_{-1}^{1}\frac{dz_{s}}{2} \; (1-z_{s}^{2}) \; F_{1}(t(s,z_{s}))\,,
\label{Eq:Fhat}
\end{align}
where the inhomogeneity $\hat{F}_{1}(s)$ contains the $s$-channel projection of the left-hand cut contributions due to the $t$- and $u$-channels, and its evaluation in the decay region requires a proper analytical continuation~\cite{Bronzan:1963mby}. 
Assuming elastic unitarity with only two-pion intermediate states, we arrive at the KT equation for the $J/\psi\to3\pi$ decay, \ie the
unitarity relation for the isobar amplitude $F_{1}(s)$:
\begin{align}
\label{Eq:discOmegaPiSingleVariable}
    {\rm{disc}}\,F_{1}(s)&=2i\left(F_{1}(s)+\hat{F}_{1}(s)\right) \; \sin\delta_{1}(s) \; e^{-i\delta_{1}(s)} \; \theta(s-4m_{\pi}^{2})\,,
\end{align}
where $\delta_{1}(s)$ is the $P$-wave $\pi\pi$ phase shift, which is real. 

Given the discontinuity relation in Eq.\,(\ref{Eq:discOmegaPiSingleVariable}), one can write an unsubtracted dispersion relation for $F_{1}(s)$ as
\begin{align}
\label{Eq:DRforF(s)}
    F_{1}(s)=\frac{1}{2\pi i}\int_{4m_{\pi}^2}^{\infty}d s' \; \frac{{\rm{disc}}\,F_{1}(s')}{s'-s}\,,
\end{align}
the solution of which can be written as:
\begin{equation}
F_{1}(s)=\Omega_{1}(s)\left(a+\frac{s}{\pi}\int_{4m_{\pi}^{2}}^{\infty}\frac{ds^{\prime}}{s'}\frac{\sin\delta_{1}(s^{\prime}) \, \hat{F}_{1}(s^{\prime})}{|\Omega_{1}(s^{\prime})|\left(s^{\prime}-s\right)}\right)\,,
\label{Eq:KTGeneral}
\end{equation}
where $\Omega_{1}(s)$ is the usual Omn\`{e}s function~\cite{Omnes:1958hv},
\begin{equation}
    \Omega_{1}(s)=\exp\left[\frac{s}{\pi}\int_{4m_{\pi}^{2}}^{\infty}\frac{ds^{\prime}}{s^{\prime}}\frac{\delta_{1}(s^{\prime})}{s^{\prime}-s}\right]\,.
\label{Eq:Omnes}
\end{equation}
The subtraction constant $a$ in Eq.~(\ref{Eq:KTGeneral}) is the only free parameter 
in the model.
It is in general complex, $a=|a|\,e^{i\phi_{a}}$. 
While its modulus $|a|$ can be fixed from the experimental $J/\psi\to3\pi$ decay width, no observable of the decay is sensitive to the overall phase $\phi_a$, so we can set $\phi_{a}=0$.  Since it determines the overall normalization of the amplitude, the constant $a$ can be factored out.

\begin{figure}
\centering\includegraphics[scale=0.45]{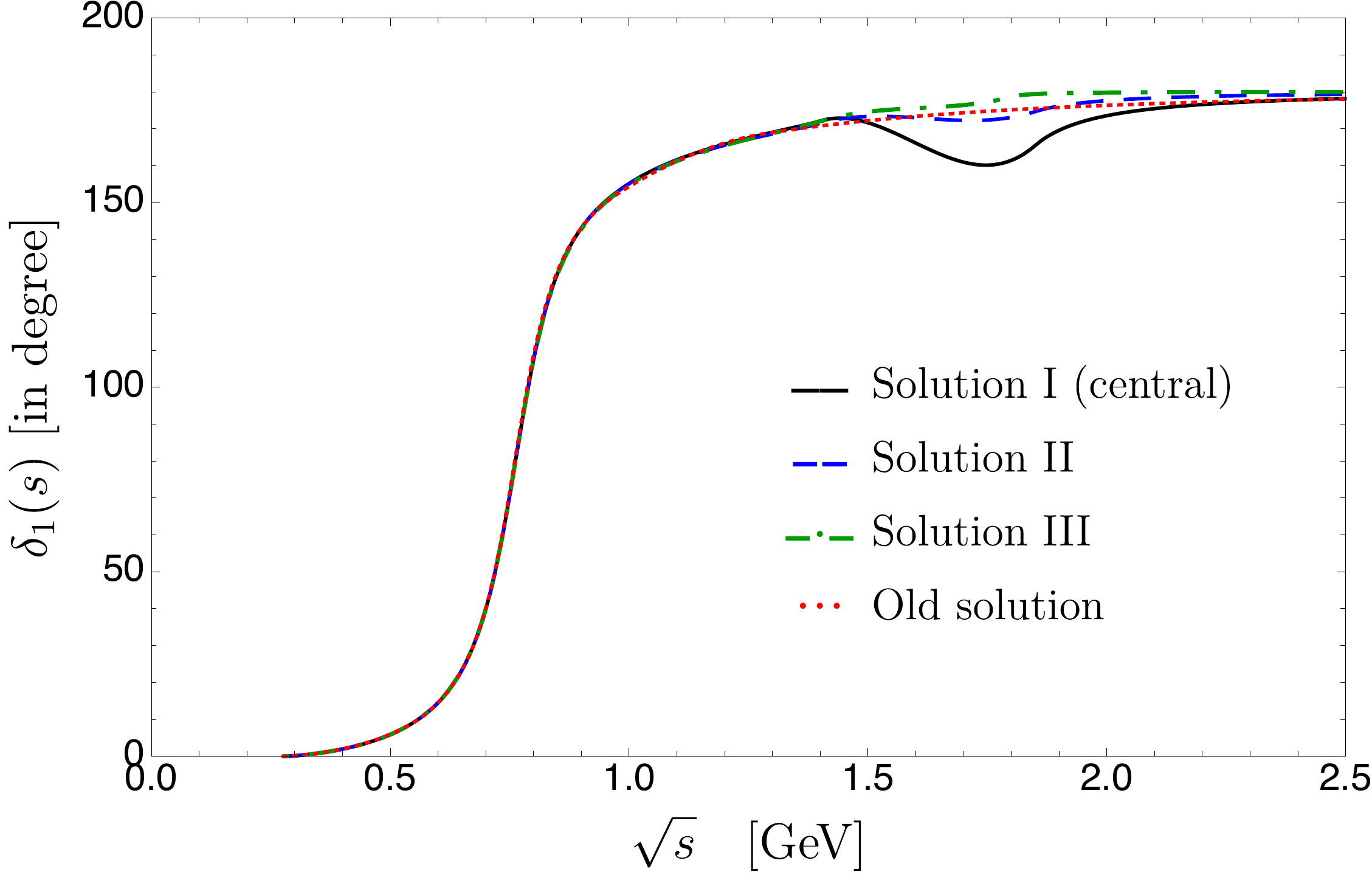}
\caption{Solutions I, II and III for the $P$-wave phase shift $\delta_{1}(s)$ from Ref.\,\cite{Pelaez:2019eqa} valid roughly up to $\sqrt{s}=2$ GeV. 
The solution of Ref.\,\cite{Garcia-Martin:2011iqs} (dotted red line) is valid only up to about $\sqrt{s}=1.3$ GeV, and is shown for completeness.}
\label{Fig:PhaseInput} 
\end{figure}

We  note that due to the asymptotic behavior of $F_{1}(s)$ in Eq.~\eqref{Eq:KTGeneral}, the amplitude $F(s,t,u)$ satisfies the Froissart-Martin bound~\cite{Froissart:1961ux,Martin:1962rt,Niecknig:2012sj}.
Also note that, even though $F_{1}(s)/\Omega_{1}(s)$ in Eq.~\eqref{Eq:KTGeneral} looks like a once-subtracted dispersion relation, $F_{1}(s)$ actually satisfies the unsubtracted dispersion relation given in Eq.~\eqref{Eq:DRforF(s)}. 
Therefore, the energy dependence of $F_{1}(s)$ is a pure prediction given solely by the phase shift $\delta_{1}(s)$. 
Here, we take $\delta_{1}(s)$ from the phase shift parametrizations of Ref.\,\cite{Pelaez:2019eqa} that are valid roughly up to $\sqrt{s}=2$ GeV.
These phase shifts contain information about inelastic channels, but given that the inelasticity is found to be rather small until about 1.4 GeV we refrain to consider them.
Therefore, the phase shift that we employ have the physics of the $\rho(770)$ and also the effects of the higher $\rho(1450)$ and $\rho(1770)$ resonances.
For our analysis, beyond $\sqrt{s}=\Lambda\equiv1.85$ GeV we smoothly guide the $\delta_{1}(s)$ to $\pi$ through~\cite{Gonzalez-Solis:2019iod,JPAC:2021rxu}

\begin{eqnarray}
\delta_{\infty}(s)\equiv\lim_{s\to\infty}\delta_{1}(s)=\pi-\frac{\alpha}{\beta+\left(s/\Lambda^{2}\right)^{3/2}}\,,
\label{Eq:PhaseExtrapolation}
\end{eqnarray}
where $\alpha$ and $\beta$ are parameters introduced so that the phase $\delta_{1}(s)$ and its first derivative $\delta^{\prime}(s)$ are continuous at $s=\Lambda^{2}$. Their explicit expressions read
\begin{eqnarray}
\alpha=\frac{3\left(\pi-\delta_{1}(\Lambda^{2})\right)^{2}}{2\Lambda^{2}\delta_{1}^{\prime}(\Lambda^{2})}\,,\quad \beta=-1+\frac{3\left(\pi-\delta_{1}(\Lambda^{2})\right)}{2\Lambda^{2}\delta_{1}^{\prime}(\Lambda^{2})}\,.
\end{eqnarray}
This ensures the expected asymptotic $1/s$ 
behavior of $\Omega_{1}(s)$.
The three phase shifts $\delta_{1}(s)$ from Ref.\,\cite{Pelaez:2019eqa} that we use as an input are shown in Fig.~\ref{Fig:PhaseInput} up to 2.5 GeV. 
The different solutions come from using different $\pi\pi$ scattering data sources.
As seen, the behavior of the phase shift solution I suggests a large interference between the $\rho^{\prime}$ and $\rho^{\prime\prime}$, with a sizable change in the phase in the region between 1.5 and 1.8 GeV, while solutions II and III looks smoother in this region.
For our analysis, we use solution I as our central input for the phase and solutions II and III to quantify the systematic uncertainties in our calculations.

\begin{figure}\centering
\includegraphics[scale=0.325]{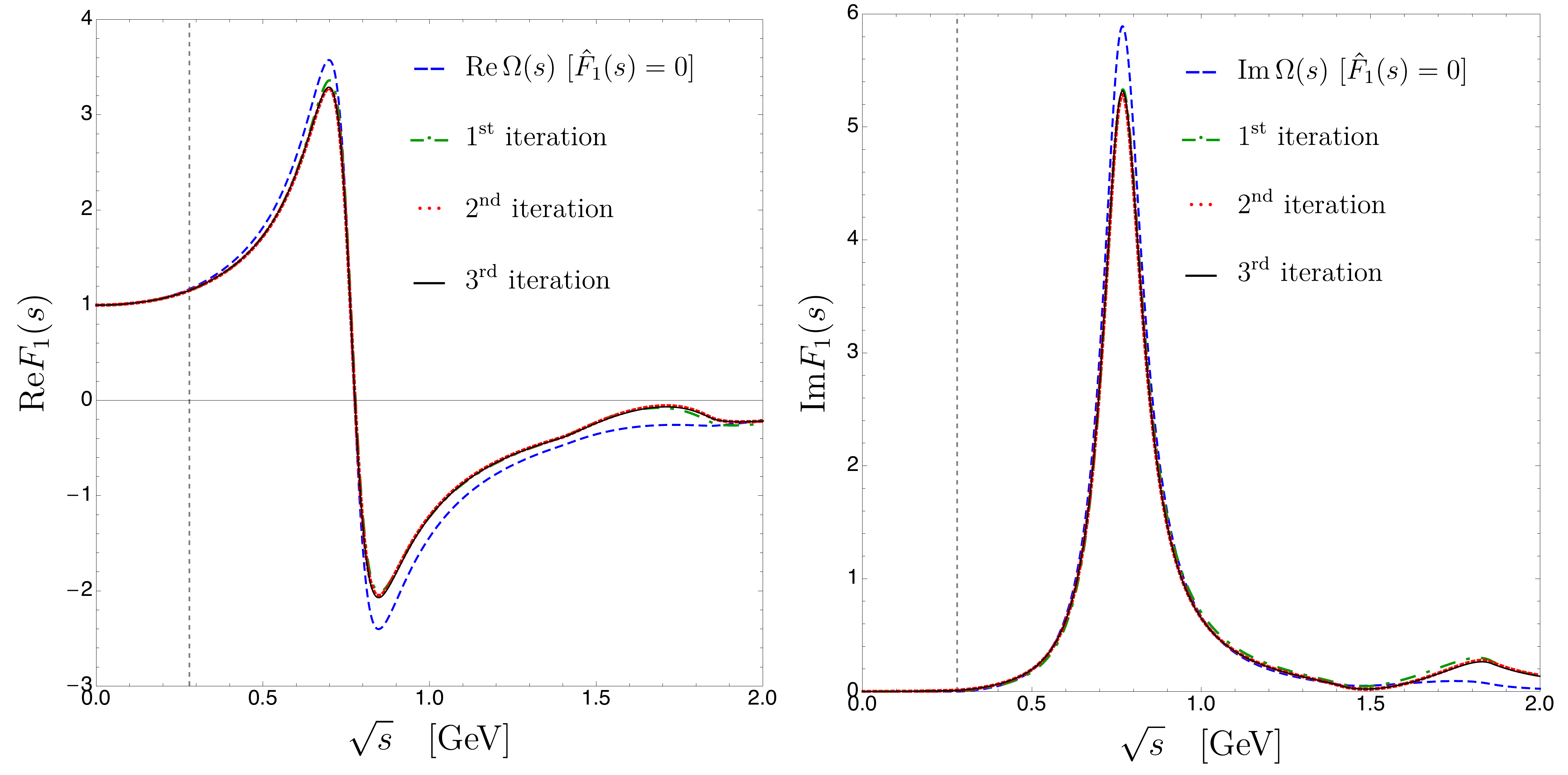}
\caption{\label{Fig:IterationsF1}Convergence behavior of the iterative procedure for the real (left plot) and imaginary (right plot) parts of the amplitude $F_{1}(s)$ given in Eq.~\eqref{Eq:KTGeneral} 
using solution I of the phase shift $\delta_{1}(s)$ as input. 
The vertical line denotes the two-pion threshold. 
}
\end{figure}

We solve Eq.~\eqref{Eq:KTGeneral} following a numerical iterative procedure similar to Refs.\,\cite{Niecknig:2012sj,Guo:2014vya,Albaladejo:2017hhj,Gasser:2018qtg,Albaladejo:2020smb}. 
We use $F_{1}(s)=\Omega_{1}(s)$ as an efficient initial input to calculate $\hat{F}_{1}(s)$ from Eq.~(\ref{Eq:Fhat}), which subsequently is inserted as an input in Eq.~\eqref{Eq:KTGeneral} for the computation of an updated $F_{1}(s)$.
This cyclic calculation is repeated until the solution converges.
In Fig.~\ref{Fig:IterationsF1}, we show the solutions for $F_{1}(s)$ (normalized to $a=1$) after each iteration step along with the initial input (dashed blue line).
As can be seen, convergence is achieved after three iterations.
The difference between the final solution (solid black) and the starting point, \ie $F_{1}(s)=\Omega_{1}(s)$ (dashed blue), is rather small,  hinting at moderate crossed-channel effects.

Note that when the crossed-channel rescattering effects are removed from the isobar $F_{1}(s)$, {\it{i.e.}} when $\hat{F}_{1}(s)=0$ in Eq.~(\ref{Eq:KTGeneral}), $F_{1}(s)$ is simply the pure Omn\`{e}s function multiplied by a constant,
\begin{equation}
    F_{1}(s)=a^{\prime}\Omega_{1}(s)\,,
    \label{Eq:OmnesIsobar}
\end{equation}
which implies the following isobar decomposition of the full amplitude (\cf~Eq.~(\ref{Eq:KTdecomposition})):
\begin{equation}
    F(s,t,u)=a^{\prime}\left(\Omega_{1}(s)+\Omega_{1}(t)+\Omega_{1}(u)\right)\,.
    \label{Eq:Omnesdecomposition}
\end{equation}
In this case, a new normalization constant $a^{\prime}$ has to be chosen to reproduce the $J/\psi\to3\pi$ decay width.
Also note that Eq.~\eqref{Eq:KTGeneral} can be written in the form 
\begin{equation}
F_{1}(s)=\Omega_{1}(s)\left(a+b'\,s+\frac{s^2}{\pi}\int_{4m_{\pi}^{2}}^{\infty}\frac{ds^{\prime}}{(s')^2}\frac{\sin\delta_{1}(s^{\prime}) \, \hat{F}_{1}(s^{\prime})}{|\Omega_{1}(s^{\prime})|\left(s^{\prime}-s\right)}\right)\,,
\label{Eq:KTGeneral_2}
\end{equation}
where $b'$ satisfies the following sum rule~\cite{Niecknig:2012sj}:
\begin{equation}\label{Eq:SumRuleAn}
b\equiv b'/a=\frac{1}{\pi}\int_{4m_{\pi}^{2}}^{\infty}\frac{ds^{\prime}}{(s')^2}\frac{\sin\delta_{1}(s^{\prime}) \,\hat{F}_{1}(s^{\prime})/a}{|\Omega_{1}(s^{\prime})|}\,.
\end{equation}
The subtraction constant, $b$, is complex due to the presence of the three-particle cut in the physical region of the decay amplitude.
This value is found to be:
\begin{equation}\label{Eq:SumRuleNum}
b_\text{sum} \simeq\, 0.141\,e^{2.321\,i}\ \text{GeV}^{-2}~.
\end{equation}
Had we used solution II or III of the phase shift $\delta_{1}(s)$ (\cf Fig.~\ref{Fig:PhaseInput}), we would have obtained $b_\text{sum} \simeq\, 0.129\,e^{2.640\,i}$ GeV$^{-2}$ and $b_\text{sum} \simeq\, 0.124\,e^{2.811\,i}$ GeV$^{-2}$, respectively. 

\begin{figure}\centering
\includegraphics[scale=0.325]{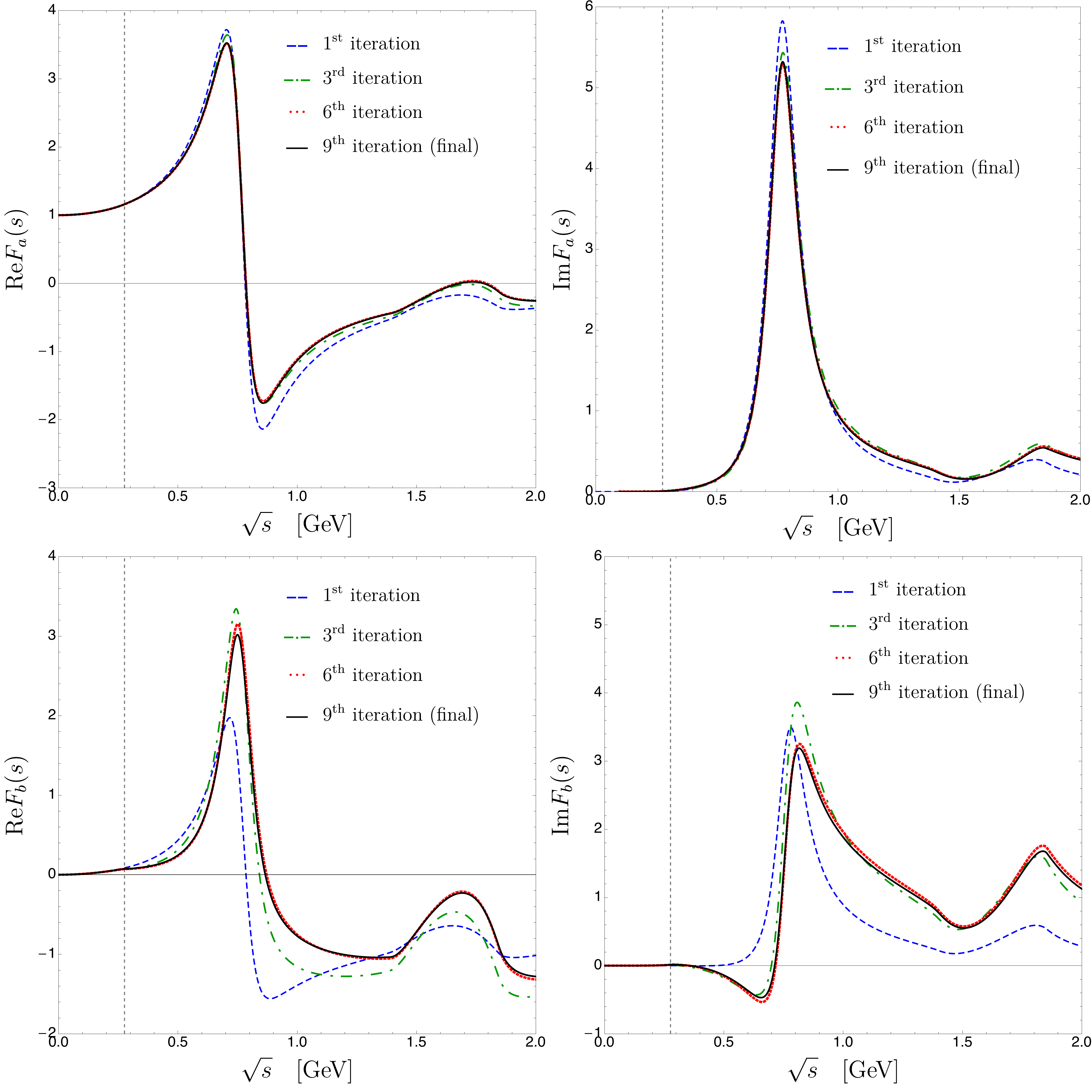}
\caption{\label{Fig:IterationsFaFb}Convergence behavior of the iterative procedure for the real (left plots) and imaginary (right plots) parts of the amplitudes $F_{a}(s)$ (Eq.~\eqref{Eq:Fa}, upper plots) and $F_{b}(s)$ (Eq.~\eqref{Eq:Fb}, lower plots) using solution I of the phase shift as $\delta_{1}(s)$ input.
The vertical line denotes the two-pion threshold.}
\end{figure}

Performing one subtraction on Eq.~\eqref{Eq:DRforF(s)} leads to the solution~\cite{Niecknig:2012sj,Albaladejo:2017hhj,Albaladejo:2019huw}:
\begin{subequations}\label{eqs:ourFs}
\begin{equation}
    F_{1}(s)=a\left[F_{a}(s)+b\,F_{b}(s)\right]\,,
\label{Eq:KTtwosub}
\end{equation}
where now $b$ is not constrained to satisfy Eq.~\eqref{Eq:SumRuleAn}, and the functions $F_{a}(s)$ and $F_{b}(s)$ are given by
\begin{eqnarray}
    F_{a}(s)&=&\Omega_{1}(s)\left[1+\frac{s^{2}}{\pi}\int_{4m_{\pi}^{2}}^{\infty}\frac{ds^{\prime}}{s^{\prime2}}\frac{\sin\delta_{1}(s^{\prime}) \, \hat{F}_{a}(s^{\prime})}{|\Omega_{1}(s^{\prime})|(s^{\prime}-s)}\right]\,,\label{Eq:Fa}\\[1ex]
    F_{b}(s)&=&\Omega_{1}(s)\left[s+\frac{s^{2}}{\pi}\int_{4m_{\pi}^{2}}^{\infty}\frac{ds^{\prime}}{s^{\prime2}}\frac{\sin\delta_{1}(s^{\prime}) \, \hat{F}_{b}(s^{\prime})}{|\Omega_{1}(s^{\prime})|(s^{\prime}-s)}\right]\,.
\label{Eq:Fb}
\end{eqnarray}
\end{subequations}
These functions only need to be calculated once since they are independent of the numerical values of $a$ and $b$ and, as we will discuss in Sec.~\ref{sec:results}, $a$ and $b$ will become fit parameters. 
In Fig.~\ref{Fig:IterationsFaFb}, we show the solutions for $F_{a}(s)$ and $ F_{b}(s)$ using a numerical iterative procedure similar to the one described previously. 
In this case, nine iterations are needed to 
obtain convergent solutions.
Strictly speaking, the amplitude $F(s,t,u)$ built from $F_{1}(s)$ in Eq.~\eqref{Eq:KTtwosub} does not satisfy the asymptotic Froissart-Martin bound
for an arbitrary value of the parameter $b \neq b_\text{sum}$ [\cf Eq.~\eqref{Eq:SumRuleNum}]. 
The main advantage of introducing one subtraction is that, due to the additional $1/s'$ factor introduced, we reduce the importance of the high energy region of the dispersion integrals
where the phase shift is not well known.
By letting the subtraction constant $b$ be a free parameter, we can partially absorb our ignorance of the higher energy part of the integral. 
This allows us to parametrize some unknown energy dependence of the $J/\psi\to 3\pi$ interaction not directly related to $\pi\pi$ rescattering.
As we will show in the following section, the once-subtracted parametrization 
provides a good representation of the data from BESIII in the $\rho(770)$ resonance region. 

\section{Results}\label{sec:results}

\subsection[$P$-wave contribution]{\boldmath $P$-wave contribution}\label{sec:resultsFwave}

\begin{figure}\centering
\includegraphics[scale=0.6]{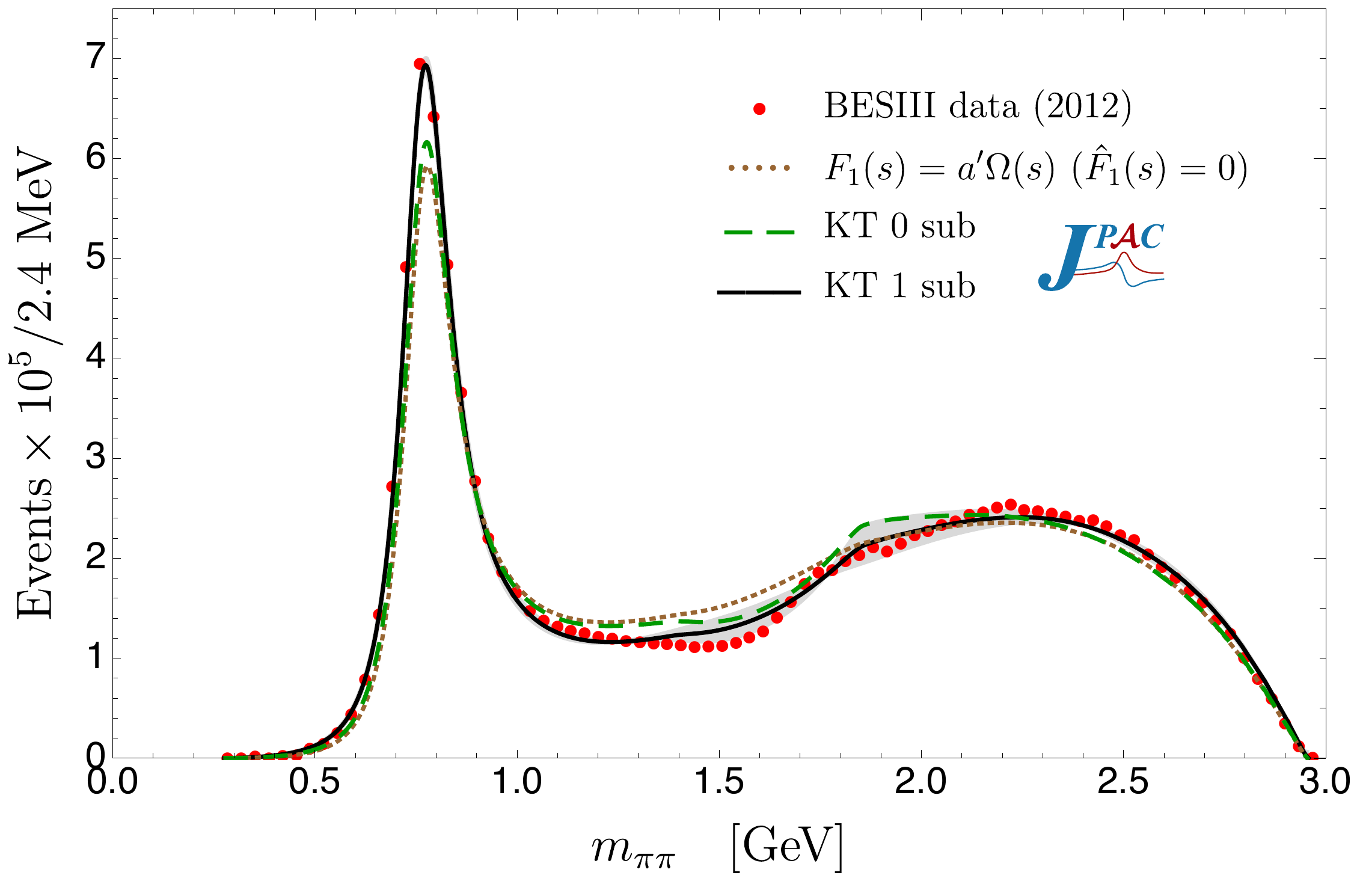}
\caption{\label{Fig:FitsToBESIII}BESIII (red circles)~\cite{BESIII:2012vmy} measurement of the $m_{\pi\pi}$ invariant mass distribution for the decay $J/\psi\to3\pi$ as compared to our prediction without crossed-channel effects (dotted brown line), with the unsubtracted KT amplitude (dashed green line) and our fit in Eq.~(\ref{Eq:FitsPwave}) including one subtraction (black solid line).
The gray band accounts for the systematic uncertainties attached to our calculations. See main text for details.}
\end{figure}

We now compare our KT amplitudes defined in the previous section to the experimental data from the BESIII collaboration~\cite{BESIII:2012vmy}.
Given that the Dalitz plot distribution is not publicly available, we are only able to analyze the di-pion mass projection of Eq.~\eqref{Eq:DecayWidth}, computed on the $\sqrt{s} \equiv m_{\pi\pi}$ invariant mass, shown in Fig.\,2 of Ref.\,\cite{BESIII:2012vmy}.
A Poisson distribution is assumed to assign uncertainty for every bin. 
High statistics of the data sample make it challenging to achieve an accurate description of the data with reasonably simple models. Nevertheless, we will be able to obtain a qualitative description of the data in the whole energy range.
We start by using the unsubtracted KT amplitude Eq.~(\ref{Eq:KTGeneral}).
The single free parameter $a$ only affects the overall normalization of the amplitude and can be fixed from the $J/\psi\to\pi^{+}\pi^{-}\pi^{0}$ decay width.
Using the PDG values $\Gamma_{J/\psi}=92.6$ keV and ${\rm{BR}}(J/\psi\to\pi^{+}\pi^{-}\pi^{0})=2.10(8)\%$~\cite{Workman:2022ynf} one finds $|a|\simeq0.051$ GeV$^{-3}$. 
In Fig.~\ref{Fig:FitsToBESIII}, we compare our prediction to the $m_{\pi\pi}$ distribution 
by BESIII with proper normalization [\cf Eq.~(\ref{Eq:Chi2})].
In the figure, we also show the result obtained when the crossed-channel rescattering is neglected [\cf~Eq.~(\ref{Eq:Omnesdecomposition})], in which case the global normalization is found to be $|a^{\prime}|\simeq0.046$ GeV$^{-3}$.
As can be observed, the result of the latter solution (dotted brown line) lies below that of the unsubtracted KT $F_{1}(s)$ solution at the peak of the $\rho$-meson, and neither reproduce the experimental data in this region.
In addition, both appear to fail at describing the intermediate energy region.
In order to achieve a better description of the data, we next use the more flexible, once-subtracted amplitude Eqs.~(\ref{Eq:Fa}) and (\ref{Eq:Fb}), with the additional subtraction constant $b$ fitted to BESIII data.
For our analysis, we define 
\begin{equation}
\chi^{2}_{{\rm{data}}}=\sum_{i=1}\left(\frac{ N_{\text{ev},i}-\mathcal{N}d\Gamma_{i}^{\rm{th}}/dm_{\pi\pi}}{\sigma_{N_{\text{ev},i}}}\right)^{2}\,,
\label{Eq:Chi2}
\end{equation}
where $N_{\text{ev},i}$ and $\sigma_{N_{\text{ev},i}}$ are, respectively, the experimental number of events distribution and the corresponding error in the $i$-th bin and
$d\Gamma_{i}^{\rm{th}}/dm_{\pi\pi}$ is the theoretical expression for the decay distribution [\cf~Eq.~(\ref{Eq:DecayWidth})].
For $\sigma_{N_{\text{ev},i}}$ we take $\sqrt{N_{{\rm{ev}},i}}$.
The constant $\mathcal{N}$ is at this stage an arbitrary normalization.
Since we are not determining the branching ratio, we reabsorb the global normalization of the amplitude $a$ into $\mathcal{N}$ and fix alone this overall constant from the fit to the BESIII data.
The sum in Eq.~(\ref{Eq:Chi2}) runs over the 80 data points and we take into account an efficiency of about $0.3$ for the number of events and the errors in our fits~\cite{BESIII:2012vmy}.

The $\chi^{2}_{{\rm{data}}}$ minimization yields 
\begin{equation}
    |b|=0.198(35)~{\rm{GeV^{-2}}}\,,\quad \phi_{b}=2.675(300)\,,
    \label{Eq:FitsPwave}
\end{equation}
which implies $|a|=0.0565(22)$ GeV$^{-3}$ for the normalization of the amplitude upon using the ${\rm{BR}}(J/\psi\to\pi^{+}\pi^{-}\pi^{0})$ from the PDG.
The statistical error is negligible and the quoted error is the theoretical systematic uncertainty attached to our calculations.
This is obtained from the absolute value of the difference between the fits performed with solutions I (central solution) and III of the phase shift $\delta_{1}^{1}(s)$ (\cf~Fig.~\ref{Fig:PhaseInput}), which gives the largest variation.
We observe that the systematic errors attached are sizable, of about $18\%$ and $11\%$ for $|b|$ and $\phi_{b}$, respectively.
We also note that this value stays close to its sum-rule prediction given in Eq.~(\ref{Eq:SumRuleNum}).
Therefore, we conclude that the pion-pion $P$-wave phase shift saturates the sum rule for the $J/\psi\to3\pi$ partial wave to about $75\%$.
This result is to be compared to similar sum rules for $\omega\to3\pi$ in Ref.\,\cite{Albaladejo:2020smb}, where the fitted value of $b$ was found to be quite different than its sum-rule $b_{\rm{sum}}$, and for $\phi\to3\pi$ in Ref.\,\cite{Niecknig:2012sj}, where it was observed that the difference between the fitted $b$ and $b_{\rm{sum}}$ was small.
The result of the fit is shown in Fig.~\ref{Fig:FitsToBESIII} with the normalization of the events distribution resulting from the fits, $\mathcal{N}=7.64(33)\times10^{8}$ in units of $(2.4\,\rm{MeV})^{-1}$.
The gray error band in the figure accounts for the systematic uncertainties associated to our fits and is defined as the (symmetrized) difference between the fit results obtained with solution I of the phase shift with respect to the ones from solution III, which give the largest difference.
It can be seen that this fit provides a satisfactory description of experimental data up to $m_{\pi\pi}\sim1$ GeV (the elastic region).
However, we obtain high values of the $\chi^{2}/$dof of about 200 but this problem is not critical.
We shall come back to discuss this point below. 
Here we stress that the once-subtracted KT amplitude is able to reproduce the $\rho(770)$ function shape and note that contributions of partial waves other than the elastic $P$-wave, which is the main one, seem to be required to describe the intermediate energy region around $m_{\pi\pi}\sim1.5$ GeV.
The next allowed partial wave is the $F$-wave, which we will include in the following subsection. 
As we will see, the inclusion of an explicit $F$-wave improves the quality of the fit.

\begin{figure}\centering
\includegraphics[scale=1.15]{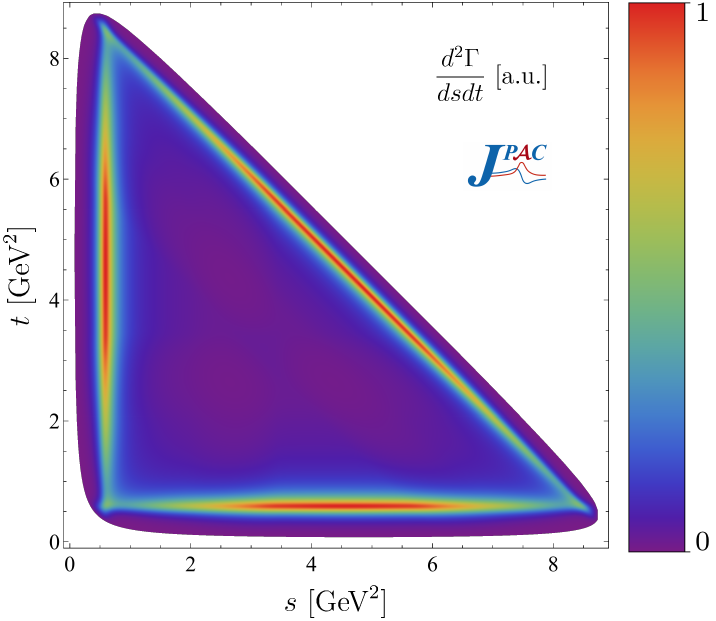}
\caption{\label{Fig:DPFitsToBESIII}Dalitz plot distribution $d^{2} \Gamma / ds \,dt$ (in arbitrary units) resulting from our fit in Eq.~(\ref{Eq:FitsPwave}).}
\end{figure}

In Fig.~\ref{Fig:DPFitsToBESIII}, we show the Dalitz plot distribution resulting from our fit, which exhibits unambiguous contributions from $\rho(770)$ resonances which appear as bands along the Dalitz plot boundaries, with almost no events in the center of the Dalitz plot.
The visual comparison with the corresponding BESIII Dalitz-plot data shows a good agreement (see Fig.~2 in Ref.\,\cite{BESIII:2012vmy}).

\subsection[Inclusion of the $F$-wave contribution]{\boldmath Inclusion of the $F$-wave contribution}\label{sec:resultsFwave}

The isobar decomposition of the amplitude including $F$-waves follows from Eq.~(\ref{Eq:AmplitudeF}) and reads~\cite{Niecknig:2012sj,Albaladejo:2020smb}:
\begin{eqnarray}\label{Eq:KTdecompositionFwave}
\begin{aligned}
F(s,t,u)&=F_1(s)+F_1(t)+F_1(u)\\[1ex]
&+(p(s)q(s))^{2}P_{3}^{\prime}(z_{s})F_3(s)+(p(t)q(t))^{2}P_{3}^{\prime}(z_{t})F_3(t)+(p(u)q(u))^{2}P_{3}^{\prime}(z_{u})F_3(u)\,,
\end{aligned}
\end{eqnarray}
where $F_{1}(s)$ is the $P$-wave isobar [\cf Eq.~(\ref{Eq:KTtwosub})], $F_{3}(s)$ is the $F$-wave isobar amplitude, which as $F_{1}(s)$ only has a right-hand cut, and:
\begin{equation}
z_{t}=\frac{s-u}{4p(t)q(t)}~, \qquad z_{u}=\frac{s-t}{4p(u)q(u)}~.
\end{equation}
The discontinuity of the $F$-wave is expressed by:
\begin{align}
\label{Eq:discFwave}
    {\rm{disc}}\,F_{3}(s)&=2i\left(F_{3}(s)+\hat{F}_{3}(s)\right) \; \sin\delta_{3}(s) \; e^{-i\delta_{3}(s)} \; \theta(s-4m_{\pi}^{2})\,,
\end{align}
where $\delta_{3}(s)$ and $\hat{F}_{3}(s)$ are the $F$-wave phase shift and inhomogeneity, respectively.
Here, we will simplify Eq.~(\ref{Eq:discFwave}) by neglecting $\hat{F}_{3}(s)$, as done for instance in Ref.\,\cite{Albaladejo:2017hhj}. 
The solution is then given by:
\begin{equation}
    F_{3}(s)=p_{3}(s)\Omega_{3}(s)\,,
\label{Eq:F3}
\end{equation}
where $\Omega_{3}(s)$ is the $F$-wave Omn\`{e}s function (\cf~Eq.~(\ref{Eq:Omnes}))
\begin{equation}
    \Omega_{3}(s)=\exp\left[\frac{s}{\pi}\int_{4m_{\pi}^{2}}^{\infty}\frac{ds^{\prime}}{s^{\prime}}\frac{\delta_{3}(s^{\prime})}{s^{\prime}-s}\right]\,.
\label{Eq:OmnesFwave}
\end{equation}

\begin{figure}
\centering\includegraphics[scale=0.3]{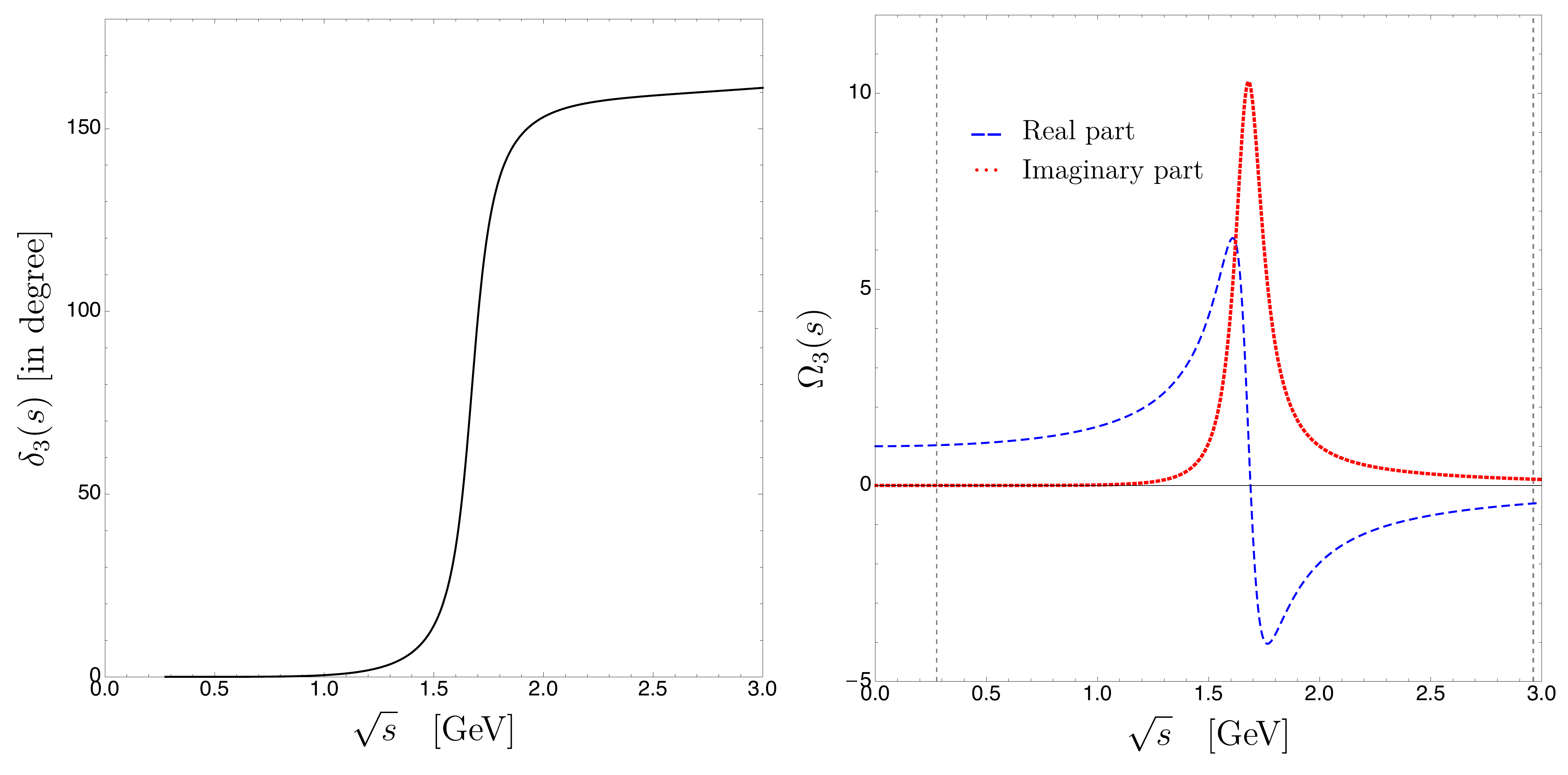}
\caption{$F$-wave phase shift $\delta_{3}(s)$ Eq.~(\ref{Eq:FwavePhase}) (left plot) and output for the Omn\`{e}s function $\Omega_{3}(s)$ Eq.~(\ref{Eq:OmnesFwave}) (right plot).}
\label{Fig:FwaveInput} 
\end{figure}

In order to obtain the required input phase $\delta_{3}(s)$, we model the $F$-wave contribution by a $\rho_{3}(1690)$ resonance ($J^{PC}=3^{--}$).
While the dominant decay mode of the $\rho_{3}(1690)$ is to $4\pi$, we only consider here its decay to $\pi\pi$ and neglect inelastic channels effects.
We use the following Breit-Wigner representation for $F_{3}(s)$:

\begin{equation}
	F_3(s)|_{\rm{BW}}=\frac{m_{\rho_{3}}^{2}}{m_{\rho_{3}}^{2}-s-im_{\rho_{3}}\Gamma^{\ell=3}_{\rho_{3}}(s)}\,,
	\label{Eq:FwaveAnsatz}
	\end{equation}
with the energy-dependent width given by
\begin{eqnarray}
	\Gamma^{\ell}_{R}(s)&=&\frac{\Gamma_{R}m_{R}}{\sqrt{s}}\left(\frac{p(s)}{p(m_{R}^{2})}\right)^{2\ell+1}\left(F_{R}^{\ell}(s)\right)^{2}\,.
\end{eqnarray}
The $F_{R}^{\ell}(s)$ denotes the Blatt-Weisskopf factor that limits the growth of the isobar~\cite{Blatt:1952ije}. 
For $\ell=3$ it is given by:
\begin{eqnarray}
	F_{R}^{\ell=3}(s)&=&\sqrt{\frac{z_{0}(z_{0}-15)^{2}+9(2z_{0}-5)^{2}}{z(z-15)^{2}+9(2z-5)^{2}}}\,,\quad
	z=r_{R}^{2}p^{2}(s)\,,\quad z_{0}=r_{R}^{2}p^{2}(m_{\rho_{3}}^{2})\,,
	\end{eqnarray}
with the hadronic scale $r_{R}=2\gev^{-1}$. The phase can then be computed from the relation
\begin{equation}
	\tan\delta_{3}(s)=\frac{{\rm{Im}}F_{3}(s)|_{\rm{BW}}}{{\rm{Re}}F_{3}(s)|_{\rm{BW}}}\,,
	\label{Eq:FwavePhase}
\end{equation}
which completes our representation of the $F$-wave isobar $F_{3}(s)$.
Using $m_{\rho_{3}}=1688$ MeV and $\Gamma_{\rho_{3}}=161$ MeV from the PDG, in Fig.~\ref{Fig:FwaveInput} we display the model for the phase $\delta_{3}(s)$ Eq.~(\ref{Eq:FwavePhase}) and the output for the corresponding Omn\`{e}s function $\Omega_{3}(s)$ Eq.~(\ref{Eq:OmnesFwave}) that we use for our analysis.

\begin{figure}\centering
\includegraphics[scale=0.6]{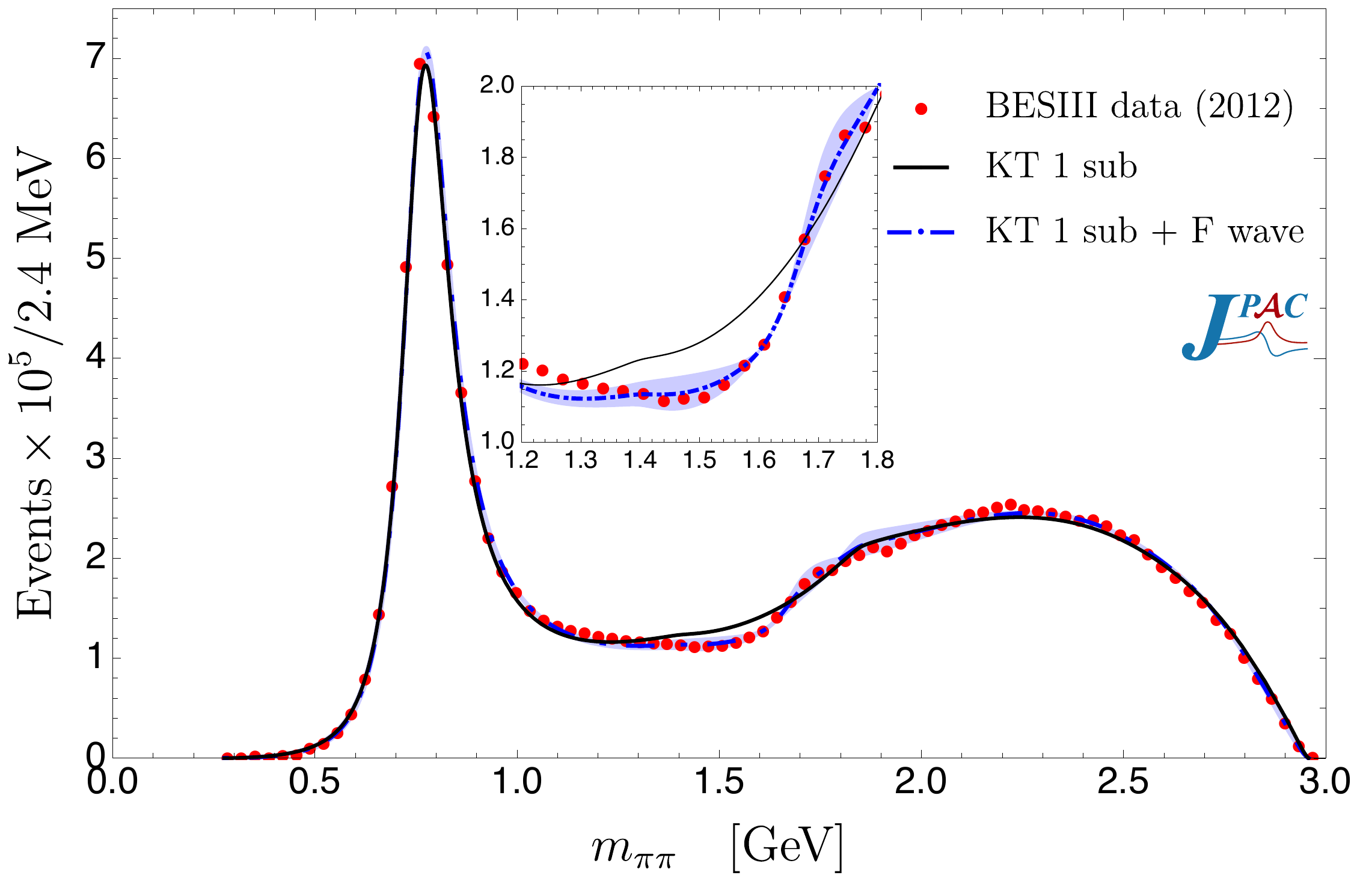}
\caption{\label{Fig:FitsToBESIIIFwave}BESIII (red circles)~\cite{BESIII:2012vmy} measurement of the $m_{\pi\pi}$ invariant mass distribution for the decay $J/\psi\to3\pi$ as compared to our fits in Eqs.~(\ref{Eq:FitsPwave}) (solid black line), (\ref{Eq:FitsFwave1}) and (\ref{Eq:FitsFwave2}) (dot-dashed blue line).
The blue error band accounts for the systematic uncertainties attached to our calculations. See main text for details.}
\end{figure}

Finally, the function $p_{3}(s)$ in Eq.~\eqref{Eq:F3} is a polynomial that parametrizes the energy dependence not directly related to the propagation of the $\rho_{3}(1690)$ resonance and fixes the strength of the $F$-wave amplitude. 
In order to achieve a satisfactory description of the data, we take $p_{3}(s)$ linear in $s$ with parameters relative to the $P$-wave amplitude, {\it{i.e.}} $p_{3}(s)=a(|c|e^{i\phi_{c}}+|d|e^{i\phi_{d}}\,s)$,
such that the overall normalization of the amplitude $a$ can be factored out in Eq.~(\ref{Eq:KTdecompositionFwave}) and absorbed in $\mathcal{N}$ (\cf Eq.~(\ref{Eq:Chi2})) as in the previous subsection.
By minimizing Eq.~(\ref{Eq:Chi2}), we obtain the following values for the fit parameters:
\begin{equation}
    |b|=0.205(34)~{\rm{GeV^{-2}}}\,,\quad \phi_{b}=2.784(298)\,,
    \label{Eq:FitsFwave1}
\end{equation}
for the $P$-wave subtraction constant, and 
\begin{eqnarray}
\begin{aligned}
    |c|\times10^{2}&=4.38(1.46)~{\rm{GeV^{-4}}}\,,\quad &\phi_{c}&=3.80(5)\,,\\[1ex]
    |d|\times10^{2}&=1.58(46)~{\rm{GeV^{-6}}}\,,\quad &\phi_{d}&=0.65(8)\,,
\end{aligned}
    \label{Eq:FitsFwave2}
\end{eqnarray}
for the parameters of the $F$-wave subtracted polynomial $p_{3}(s)$.
Again, the quoted error in the previous equations is the systematic uncertainty obtained from using the different $P$-wave phase shifts $\delta_{1}(s)$ as input.
The result of this fit implies $|a|=0.0581(60)$ GeV$^{-3}$ for the overall normalization of the amplitude and it is plotted in Fig.~\ref{Fig:FitsToBESIIIFwave} as the dash-dotted blue line using the event distribution normalization from the fits, $\mathcal{N}=8.09(41)\times10^{8}$ in units of $(2.4\,\rm{MeV})^{-1}$.
In the figure, the result of the standalone $P$-wave fit [\cf~Eq.(\ref{Eq:FitsPwave})] is also shown for comparison.
As seen, the $\rho_{3}(1690)$-induced $F$-wave contribution improves the description of the data around 1.5 GeV.
Numerically, we find that the individual $F$-wave contribution is rather small, while the interference between the $P$- and $F$-waves gives a correction of a few percent in the region $m_{\pi\pi}\sim1.5$ GeV.
The $\chi^{2}/$dof remains high (about 100).
However, with the systematic uncertainties associated to our fits (blue error band in Fig.~\ref{Fig:FitsToBESIIIFwave}),
we conclude that our representation of the amplitude is capable of describing the two more prominent features shown by the data: the line shape of the BESIII measurements in the vicinity of the $\rho(770)$ resonance as well as the movement of the function at $m_{\pi\pi}\sim1.5$ GeV due to the $F$-wave effects.\footnote{We shall wait for the arrival of new Dalitz distribution experimental data from BESIII to ascribe a strict statistical meaning to our $\chi^{2}$ fits.}
As for the Dalitz-plot distribution, the $F$-wave effects provides no significant change with respect to Fig.~\ref{Eq:FitsPwave} and we thus refrain to show them here.
\section{\boldmath $J/\psi\to\pi^{0}\gamma^{*}$ transition form factor}\label{sec:TFF}

\begin{figure}\centering
\includegraphics[width=5cm]{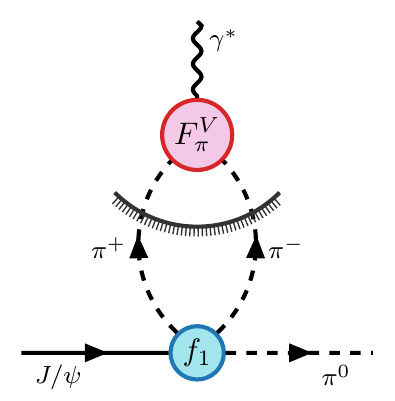}
\caption{Diagrammatic representation of the two-pion contribution to the discontinuity of the $J/\psi\pi^{0}$ transition form factor [\cf Eq.\,(\ref{Eq:Discontinuity})].
The blue and red circles represent, respectively, the full $s$-channel $P$-wave $J/\psi\to3\pi$ amplitude $f_{1}(s)$ and the pion vector form factor $F_{\pi}^{V}(s)$.\label{fig:tff}}
\end{figure}

The $J/\psi\pi^0$ transition form factor (TFF), $f_{J/\psi\pi^{0}}(s)$, governs the $J/\psi\to \pi^0 \gamma^{\ast}$ amplitude and its energy dependence is experimentally accessible from the 
decays $J/\psi\to\pi^{0}e^{+}e^{-}$ and $J/\psi\to\pi^{0}\mu^{+}\mu^{-}$. 
At present, there is no measurement of the shape of the form factor and the only experimental information on these decays is the measurement of the branching ratio by the BESIII collaboration, $BR(J/\psi\to\pi^{0}e^{+}e^{-})=(7.56\pm1.32\pm0.50)\times10^{-7}$~\cite{BESIII:2014dax}.
This measurement was obtained subtracting the $\rho$ resonance contribution and assuming that excited $c\bar{c}$ exchanges, {\it{e.g.}} coming from off-shell $\psi'$ contributions, dominate the energy-dependence of the form factor. 
Refs.\,\cite{Chen:2014yta,Kubis:2014gka} showed that subtracting this contribution is not well motivated, as the light vector meson contributions to the form factor actually dominate the decay.   
Using the formalism previously employed for the decays of light vector mesons $\omega/\phi\to\pi^{0}\gamma^{*}$~\cite{Schneider:2012ez,JPAC:2020umo}, we present a dispersive description of $f_{J/\psi\pi^{0}}(s)$ comparable to Ref.\,\cite{Kubis:2014gka}, but with the difference that our analysis is driven by the $J/\psi\to3\pi$ experimental data analysis presented in Sec.~\ref{sec:results}.

A dispersive representation of $f_{J/\psi\pi^{0}}(s)$ is fully determined, up to possible subtractions, by the discontinuity across the right hand cut. 
Here, we focus on the light-quark resonance contributions to the discontinuity, which dominate the form factor at low and intermediate energies.
Additional $c\bar{c}$ contributions can arise close to the upper limit of the accessible phase space, $\sqrt{s}=m_{J/\psi}-m_{\pi^{0}}$, and in fact can dominate the transition form factor there~\cite{Chen:2014yta,Kubis:2014gka}, but these
contributions appear in a region of the Dalitz decays which are strongly suppressed by phase space~\cite{Chen:2014yta,Kubis:2014gka}, rendering the task of experimentally observing them nearly impossible.
Bearing this in mind, and because of the absence of experimental data for the form factor, we do not consider them in our analysis.

In order to be consistent with the elastic approximation in the $J/\psi \to \pi^{+}\pi^{-}\pi^{0}$ study, we include only the two-pion intermediate state contribution to the discontinuity (see Fig.~\ref{fig:tff} for a diagrammatic  interpretation):
\begin{equation}
{\rm{disc}}f_{J/\psi\pi^{0}}(s)=i \; \frac{p^{3}(s)}{6\pi\sqrt{s}} \; {F_{\pi}^{V}}^{*}(s) \; f_{1}(s) \; \theta(s-4m_{\pi}^{2})\,,
\label{Eq:Discontinuity}
\end{equation}
which requires as input the full $s$-channel $P$-wave $J/\psi\to3\pi$ amplitude $f_{1}(s)$ given in Eq.~\eqref{Eq:f1} and 
the pion vector form factor complex-conjugate ${F_{\pi}^{V}}^{*}(s)$, which we approximate by the Omn\`{e}s function (complex-conjugate) given in Eq.~\eqref{Eq:Omnes}. 
Given that we are using a once-subtracted dispersion relation for the $J/\psi\to3\pi$ KT equations, an unsubtracted dispersion relation for the TFF,
as used for instance in Ref.\,\cite{Kubis:2014gka}, would result in a divergent integral if no cutoff is used. 
Therefore, we use a once-subtracted dispersion relation for the TFF itself,
\begin{eqnarray}
f_{J/\psi\pi^0}(s)=|f_{J/\psi\pi^0}(0)|\,e^{i\phi_{J/\psi\pi^0}(0)}+\frac{s}{12\pi^{2}}\int_{4m_{\pi}^{2}}^{\infty}\frac{ds^{\prime}}{(s^{\prime})^{3/2}}\frac{p^{3}(s^{\prime}) \; {F_{\pi}^{V}}^*(s^{\prime}) \; f_{1}(s^{\prime})}{(s^{\prime}-s)}\,,
\label{Eq:JpsiPiFF1sub}
\end{eqnarray}
where we indicate explicitly the existence of a non-vanishing phase of $f_{J/\psi\pi^0}(s)$ at $s=0$. 
This is implied by the cross-channel effects, \ie the functions ${F_{\pi}^{V}}^{*}(s)$ and $f_{1}(s)$ do not have the same phase, and the discontinuity of $f_{J/\psi\pi^0}(s)$ is in general complex~\cite{Schneider:2012ez,JPAC:2020umo}. 
The modulus of the subtraction constant  $|f_{J/\psi\pi^0}(0)|$ can be fixed from the $J/\psi\to\pi^{0}\gamma$ partial decay width 
\begin{equation}
\Gamma(J/\psi\to\pi^{0}\gamma)=\frac{e^{2}(m_{J/\psi}^{2}-m_{\pi^{0}}^{2})^{3}}{96\pi m_{J/\psi}^{3}}\; |f_{J/\psi\pi^{0}}(0)|^{2}\,.
\label{Eq:JpsiPiGammaWidth}
\end{equation}
Using the value of the partial decay width of $J/\psi\to\pi^{0}\gamma$~\cite{Workman:2022ynf} in combination with the above equation, one obtains:
\begin{equation}
|f_{J/\psi\pi^{0}}(0)|=6.0(3)\times10^{-4}\quad\rm{GeV}^{-1}\,.
\label{Eq:fJpsi0}
\end{equation}
The phase $\phi_{J/\psi\pi^{0}}(0)$ is a free parameter that can only be accessed from the transition form factor experimental data (see \eg Ref.\,\cite{JPAC:2020umo}). 
Due to the absence of data for $J/\psi\to \pi^0 \gamma^{\ast}$, we set $\phi_{J/\psi\pi^{0}}(0)=0$ in our study.

\begin{figure}\centering
\includegraphics[scale=0.6]{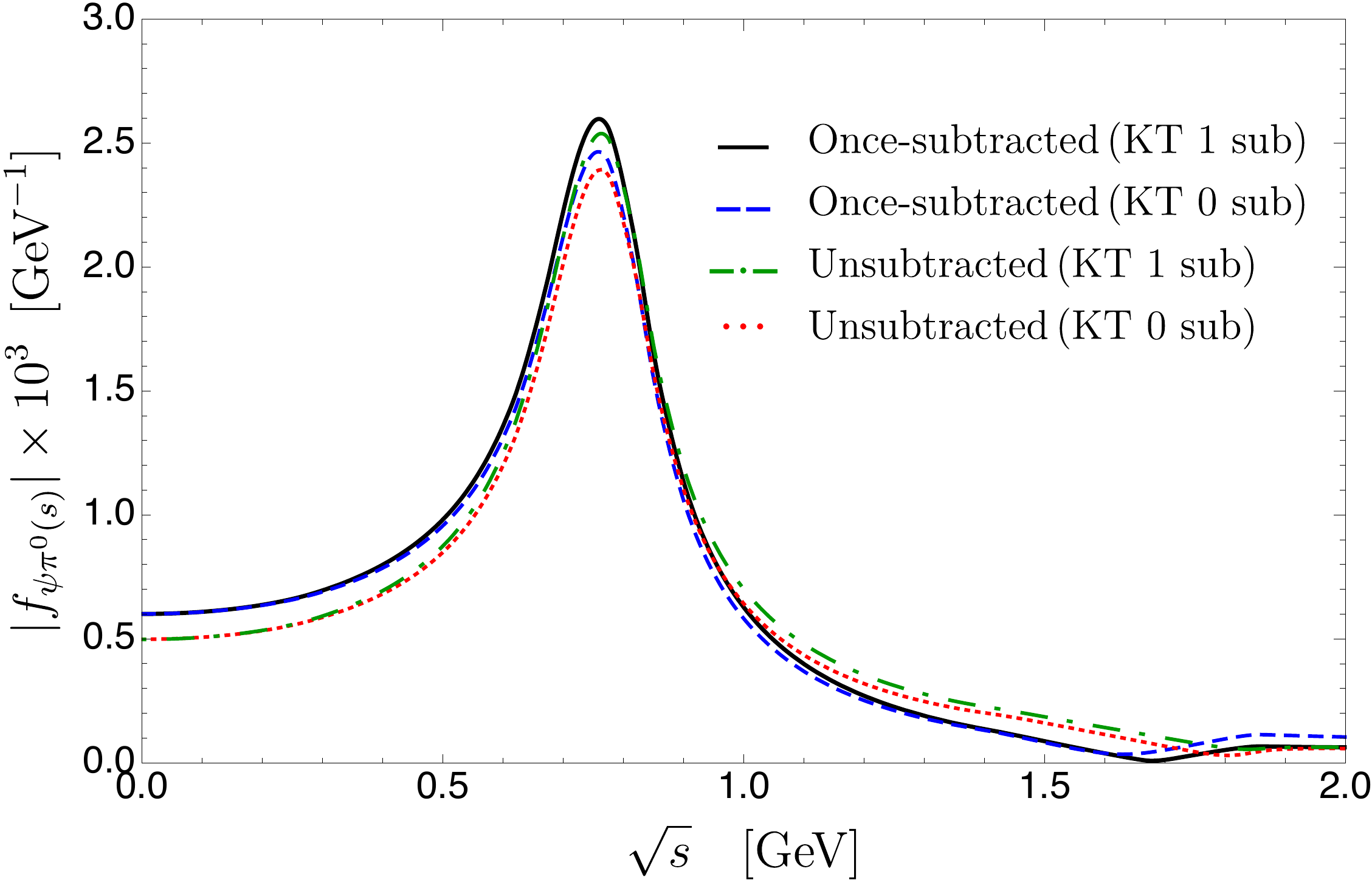}
\caption{\label{Fig:TransitionFormFactor}Prediction for the absolute value of the transition form factor $J/\psi\to\pi^{0}\gamma^{*}$ using Eq.~(\ref{Eq:JpsiPiFF1sub}) (solid black line) and variants of it. See main text for details.
}
\end{figure}

In Fig.~\ref{Fig:TransitionFormFactor}, we show up to $\sqrt{s}=2$ GeV our prediction for the absolute value of the transition form factor resulting from Eq.~(\ref{Eq:JpsiPiFF1sub}) and using the results from Eq.~(\ref{Eq:FitsPwave}) (solid black line).
This is our central result for the form factor.
In this figure, however, we also show the result of using the unsubtracted KT solution for $J/\psi\to3\pi$ (dashed blue line).
It is worth noting that both curves are similar and only a slight difference is observed at the $\rho$ peak.
Additionally, the calculations when an unsubtracted dispersion relation for the form factor is used are also shown in the figure, both with an unsubtracted (dotted red line) and once-subtracted (dot-dashed green line) $J/\psi\to3\pi$ amplitude.
In the latter case, we have cut the dispersive integral at 4 GeV$^{2}$ to avoid the dispersion relation to diverge.
Again, both curves are similar.
In this case, the value at the real photon energy can be calculated from the sum rule~\cite{Schneider:2012ez,Kubis:2014gka}:
\begin{equation}
    f_{J/\psi\pi^{0}}(0)=\frac{1}{12\pi^{2}}\int_{4m_{\pi}^{2}}^{\infty}ds^{\prime}\frac{p^{3}(s^{\prime})F_{\pi}^{V*}(s^{\prime}) f_{1}(s^{\prime})}{(s^{\prime})^{3/2}}\,.
\end{equation}
This value is found to be $|f_{J/\psi\pi^{0}}(0)|=5.0(2)\times10^{-4}$ GeV$^{-1}$
for both versions of the unsubtracted dispersion relation.
The quoted uncertainty is the systematic error from using the different phase shifts as input.
This value is in qualitative agreement with the value extracted from the measured $J/\psi\to\pi^{0}\gamma$ in Eq.~(\ref{Eq:fJpsi0}), indicating that the normalization is saturated by the two-pion intermediate state contribution by roughly $85\%$.
The difference between the various lines provides an estimate of the theoretical uncertainty associated to our description.
We expect our study to strengthen the case for new experimental measurements of the shape of this form factor, which would allow improving the understanding of radiative $J/\psi$ decays.
\section{Summary}\label{sec:conclusions}

We have analyzed the decay $J/\psi\to\pi^{+}\pi^{-}\pi^{0}$ within the framework of the Khuri-Treiman equations, which satisfy the constraints imposed by unitarity, analyticity and crossing symmetry.
We have included the $P$-wave effects of the $\pi\pi$ subsystem up to around 2 GeV, which are controlled by the $\pi\pi$ $P$-wave scattering-phase shift.
We have seen that one subtraction in the $P$-wave amplitude is necessary to achieve a good description of the experimental data in the $\rho(770)$-region.
The corresponding subtraction constant was fixed from fits to the di-pion invariant mass distribution from BESIII.
We have also seen that the $P$-wave alone is not capable of reproducing the data in the mass region around $m_{\pi\pi}\sim1.5$ GeV, and that the inclusion of
an $F$-wave contribution arising from the $\rho_{3}(1690)$ brings theory closer to data in this region.
In addition, we have provided predictions for the transition form factor $J/\psi\to\pi^{0}\gamma^{*}$ up to 2 GeV.
Our study lays the groundwork for an event-by-event likelihood fit of high-precision data from
$J/\psi$ decays, which are expected to be available from BESIII in a near future.

\acknowledgments

The authors would like to thank Joshua Jackson and Ryan Mitchell (Indiana University) for fruitful discussions.
MA is supported by Generalitat Valenciana under Grant No.~CIDEGENT/2020/002, and by the Spanish Ministerio de Ciencia e Innovación (MICINN) under contracts No.~PID2020-112777GBI00.
The work of SGS is supported by the Laboratory Directed Research and Development program of Los Alamos National Laboratory under project number 20210944PRD2, and by the U.S. Department of Energy through the Los Alamos National Laboratory. 
Los Alamos National Laboratory is operated by Triad National Security, LLC, for the National Nuclear Security Administration of U.S. Department of Energy (Contract No. 89233218CNA000001).
This work was supported by the U.S. Department of Energy contract DE-AC05-06OR23177, under which Jefferson Science Associates, LLC operates Jefferson Lab, 
U.S. Department of Energy Grants
No.~DE-FG02-87ER40365 and No.~DE-FG02-92ER40735,
CONACYT (Mexico) Grant No.~A1-S-21389, and
Spanish national Grants PID2020-118758GB-I00 
and PID2019–106080 GB-C21.
CFR is supported by Spanish Ministerio de Educaci\'on y Formaci\'on Profesional under Grant No.~BG20/00133.
VM is a Serra Húnter fellow.
The work of MM is funded by the Deutsche Forschungsgemeinschaft under Germany's Excellence Strategy-EXC-2094-390783311. DW is supported by National Natural Science Foundation of China Grant No. 12035007 and the
NSFC and the Deutsche Forschungsgemeinschaft (DFG,
German Research Foundation) through the funds provided to the Sino-German Collaborative Research Center
TRR110 “Symmetries and the Emergence of Structure
in QCD” (NSFC Grant No. 12070131001, DFG ProjectID 196253076-TRR 110).
This work contributes to the aims of the U.S. Department of Energy ExoHad Topical Collaboration, contract DE-SC0023598.



\bibliographystyle{apsrev4-1_MOD}
\bibliography{refs.bib}

\end{document}